\documentclass[table]{IEEEtran}

\usepackage{amssymb}
\usepackage{amsthm}
\usepackage{authblk}
\usepackage{booktabs}
\usepackage{eurosym}
\usepackage{float}
\restylefloat{table}
\usepackage{listliketab}
\usepackage{listings}
\makeatletter
\def\lst@makecaption{%
  \def\@captype{table}%
  \@makecaption
}
\makeatother
\usepackage{multirow}
\usepackage{tabularx}

\usepackage{amsmath}
\usepackage{slashbox}
\usepackage{graphicx}
\usepackage[hyphens]{url}
\usepackage[hidelinks]{hyperref}
\usepackage{hyphenat}
\usepackage[backend=biber,sorting=none]{biblatex}
\usepackage{pifont}

\usepackage{csquotes}
\usepackage[capitalise]{cleveref}
\usepackage[shortcuts]{extdash}
\usepackage[english]{babel}
\usepackage{xcolor}
\usepackage{orcidlink}

\lstdefinelanguage{JavaScript}{
  keywords={typeof, new, true, false, catch, function, return, null, catch, switch, var, if, in, while, do, else, case, break},
  keywordstyle=\color{blue}\bfseries,
  ndkeywords={class, export, boolean, throw, implements, import, this},
  ndkeywordstyle=\color{darkgray}\bfseries,
  identifierstyle=\color{black},
  sensitive=false,
  comment=[l]{//},
  morecomment=[s]{/*}{*/},
  commentstyle=\color{purple}\ttfamily,
  stringstyle=\color{red}\ttfamily,
  morestring=[b]',
  morestring=[b]"
}

\title{Risk Assessment Graphs: Utilizing Attack Graphs for Risk Assessment}

\author{
Simon Unger $^{1}$\orcidlink{0000-0003-2379-4723},
Ektor Arzoglou $^{1}$\orcidlink{0000-0001-8664-1885},
Markus Heinrich $^{2}$\orcidlink{0000-0001-8151-2734},
Dirk Scheuermann $^{3}$\orcidlink{0000-0003-2883-0519} and
Stefan Katzenbeisser $^{1,2}$}

\affil{%
$^{1}$ \quad University of Passau; \{firstname.lastname\}@uni-passau.de\\
$^{2}$ \quad INCYDE GmbH; \{firstname.lastname\}@incyde.com\\
$^{3}$ \quad Fraunhofer SIT; dirk.scheuermann@sit.fraunhofer.de}

\newif\ifcomments
\commentstrue

\definecolor{simon}{HTML}{800000}
\definecolor{ektor}{HTML}{000088}
\definecolor{markus}{HTML}{008000}
\definecolor{dirk}{HTML}{0000FF}

\newcommand{\mycomment}[3]{    
    \ifcomments
        {%
            \noindent%
            \ifhmode%
                \unskip%
            \fi%
            \color{#1}%
            \textbf{\scriptsize{#2:}} #3%
        }%
    \fi
}

\usepackage[breakable,most]{tcolorbox}
\tcbset{
    breakable,
    title=Notes,
    enhanced
}
\usepackage{comment}
\ifcomments
    
\else
    \excludecomment{mybox}
\fi

\begin{document}

\maketitle


\begin{abstract}
    Risk assessment plays a crucial role in ensuring the security and resilience of modern computer systems. 
    Existing methods for conducting risk assessments often suffer from tedious and time-consuming processes, making it challenging to maintain a comprehensive overview of potential security issues. 
    In this paper, we propose a novel approach that leverages attack graphs to enhance the efficiency and effectiveness of risk assessment. 
    Attack graphs visually represent the various attack paths that adversaries can exploit within a system, enabling a systematic exploration of potential vulnerabilities. 
    By extending attack graphs with capabilities to include countermeasures and consequences, they can be leveraged to constitute the complete risk assessment process.
    Our method offers a more streamlined and comprehensive analysis of system vulnerabilities, where system changes, or environment changes can easily be adapted and the issues exposing the highest risk can easily be identified.
    We demonstrate the effectiveness of our approach through a case study, as well as the applicability by combining existing risk assessment standards with our method.
    Our work aims to bridge the gap between risk assessment practices and evolving threat landscapes, offering an improved methodology for managing and mitigating risks in modern computer systems.
\end{abstract}


\section{Introduction}\label{sec: intro}

Traditional cities are becoming smarter. 
One of the core smart city concepts is smart mobility, which has attracted considerable attention from security researchers due to the emergence of smart vehicles and V2X communication that have given rise to novel cybersecurity threats.

Over the last decade, several trends have contributed to the automotive and railway threat landscape. 
First, sophisticated features in smart vehicles come with a higher volume of lines of code, aggravating testability and auditing and increasing the likelihood and severity of vulnerabilities. 
Second, (wireless) communication interfaces in smart vehicles come with a higher volume of external peripheral devices that can connect to smart vehicles, hence increasing the attackers' access point options, and also with a higher volume of connections, hence increasing the risk of malicious interactions. 
Finally, a higher volume of connections between smart vehicles comes with a higher volume of exchanged data, which in most cases is personal and, therefore, immensely valuable. In other words, more data is generated and needs to be considered and protected.

Graphical security modeling is a widely-used and well-established approach for representing and analyzing threat landscapes that examine vulnerabilities of systems and organizations. 
One of the primary strengths of graphical security models is that they allow for the inclusion of user-friendly visual elements with formal semantics and algorithms, enabling both qualitative and quantitative analyses. 
Over the last couple of decades, security researchers have been progressively focusing on graphical security modeling, which has gradually evolved into a valuable tool for the assessment of risks in real-life systems, such as automotive and railway environments.

Threat landscapes include (1) malicious actions of an attacker, whose goal is to harm or damage one or more assets of a system or organization, and (2) countermeasures for either preventing or mitigating such malicious actions. 
The first \emph{tree-based approach} for graphical security modeling was the \emph{threat logic trees}, which was introduced by Weiss in 1991~\cite{weiss1991}, thereby motivating the development of several subsequent frameworks, such as attack trees, which are still considered one of the most important and favored tools for the assessment of risks to date.

In all tree-based approaches, the modeling process begins with identifying a feared event, which is shown as a root node, and continues with the refinement of the attack steps, resulting in a tree model.
However, tree structures are limited to only one path between a pair of nodes. 
In other words, with tree structures, each refined node can only have one parent node. 
This limitation is addressed by the \emph{directed acyclic graph (DAG) structure}, which enables refined nodes to have multiple parent nodes. 
As a result, DAG structures can provide a higher level of detail, but they can also come with a higher level of complexity, which can nevertheless be dealt with modularization, thereby allowing the model to be subdivided into loosely-coupled, independent, and interchangeable parts that can be studied individually and in parallel. 
Finally, while the one-to-many relationship between nodes in tree structures results in a linear analysis of the threat landscape, the many-to-many relationship between nodes in DAG structures can theoretically result in an exponential analysis.
However, the complexity is kept small in practice due to the acyclic structure, and the threat landscape analysis is eventually possible.

Ensuring the security of systems is not a static process that is over after going through once.
The conditions are constantly changing, on the one hand attackers and their capabilities are evolving, and on the other hand, systems themselves are being extended and evolving.
To effectively perform the necessary continuous security management, it is necessary to know not just the threat landscape but to be able to understand the consequences and impacts if attacks are performed successfully.
Hence, it is necessary to continuously perform a risk analysis to identify the potential exposure.
Nowadays, risk management is primarily done using large tables filled with a lot of information and use cases.
Large tables only offer limited visibility, as it is challenging to maintain a comprehensive overview of risks.
With numerous rows and columns, it becomes difficult to identify trends and patterns or prioritize risks effectively.
Furthermore, managing risk can be a tedious and time-consuming process.
Updating and maintaining tables with evolving risks and mitigation measures can require significant effort, especially when dealing with a complex system or multiple risk factors.
This gets even harder when dealing with large tables that often fail to provide the necessary context and connections between different risks.
Additionally, analyzing and interpreting data from large tables can be daunting. 
It may require specialized tools or skills to extract meaningful insights from the extensive amount of information presented in the table format.
Large tables may further lack the flexibility to accommodate changing risk scenarios or evolving requirements. 
Modifying or updating the table structure to incorporate new risks or factors can be cumbersome and may hinder agility in risk management.
With numerous cells and data entries, there is also an increased risk of errors, inaccuracies, or inconsistencies in the large table. 
These issues can undermine the reliability and integrity of the risk management process.

We propose a graphical solution for the risk management process to mitigate these disadvantages of tables.
A visual representation can enhance the understanding and communication of complex risk information and make it easier to identify patterns, trends, and relationships among risks, facilitating effective decision-making.
Complex risk data is further simplified by presenting it in a clear and concise manner.
Understanding  the relationships, dependencies, and interactions between various risk elements is necessary to understand the overall risk landscape.
Visual representations of the entire risk landscape provide this overview, allowing for the identification of interdependencies, hotspots, or areas of high vulnerability.
Graphical solutions can also aid in developing and evaluating risk mitigation strategies. 
By visually representing the potential consequences and effectiveness of different mitigation measures, decision-makers can make more informed choices and allocate resources more efficiently.
Furthermore, it allows for the exploration of different risk scenarios. 
By manipulating variables or parameters within the visual representation, it becomes possible to assess the potential impact of various risk factors and evaluate the effectiveness of different response strategies.
Additionally, as graphical solutions can be more adaptable to changing requirements and evolving risks, they allow for easier updates and modifications, enabling risk management processes to be more responsive and agile.

Consequently, we believe a graphical solution for the risk assessment process improves the maintenance of risk scenarios and facilitates accessibility to different stakeholders, including non-technical audiences.
However, the existing graphical solutions are momentarily used to describe the threat landscape.
Which, of course, is helpful for the risk management process but not sufficient to represent the entire risk management process.
Therefore, motivating us to define a new graphical method for risk assessment by extending existing graphical methods for depicting the threat landscape.
Besides ways to depict attack vectors, their probability, and countermeasures, our method includes a way to depict the consequences of attack vectors and the impact level, enabling us to calculate a risk value.

The remainder of the paper is structured as follows:
After the introduction,~\cref{sec: related work} discusses the related work.
Our definition of attack graphs is given in~\cref{sec: attack graphs}.
The necessary adjustments to use these attack graphs are presented in~\cref{sec: attack graph risk assessment}, including an example of how the risk assessment is performed in our project.
\cref{sec: applicability of attack graphs to risk management standards} validates our defined method by combing it with the risk assessment processes of ISO/SAE 21434~\cite{21434} and CLC/TS 50701~\cite{50701} respectively.
The scalability and practicality are evaluated in~\cref{sec: evaluation}.
Finally,~\cref{sec: conclusion} concludes this paper.

\section{Related Work}\label{sec: related work}

Kordy et al.~\cite{DAGpaper} categorize thirty-three frameworks for graphical analysis of attack and defense scenarios into (1) \emph{attack and/or defense modeling}, which focus on the formal aspects of attacks or defenses, and (2) \emph{static or sequential modeling}, which focus on the temporal aspects or dependencies between actions. 
Using the same categorization, this section provides an overview of all the frameworks, and it describes these frameworks that fulfill the majority of properties incorporated in the framework of this article.

By reviewing frameworks from current literature, we identify seven properties for graphically modeling and managing an entire risk landscape.
The first property is \emph{attack vectors}, which enables the relations (shown as edges) between attack steps (shown as nodes) and, therefore, the formation of attack paths (i.e., attack vectors). 
The second property is the \emph{directed acyclic graph (DAG) structure}, thereby enabling linear (i.e., directed) and finite (i.e., acyclic) series of attack steps towards multiple potential attack goals (i.e., graph). 
The third property is \emph{node attributes}, which enables the quantification and, therefore, the evaluation of attack steps. 
The fourth property is \emph{dynamic connectors}, thereby enabling extensive attack refinements (besides the basic AND-OR refinements). 
The fifth property is \emph{edge attributes}, which enables the quantification and, therefore, the evaluation of relations between attack steps. 
The sixth property is \emph{countermeasure nodes}, thereby enabling actions to reduce the negative consequences of attacks.
The final property is \emph{consequence nodes}, enabling the presentation of consequences of successful attacks, which is also necessary to constitute the impact.

\begin{table*}[h]
\rowcolors{2}{gray!10}{gray!40}
\renewcommand{\arraystretch}{1.2}
\caption{Static attack modeling frameworks compared to the seven defined properties.}
\label{tab: static attack modeling}
\noindent\makebox[\textwidth]{%
\begin{tabular}[t]{>{\raggedright}p{0.15\textwidth}>{\raggedright}p{0.06\textwidth}>{\raggedright}p{0.08\textwidth}>{\raggedright}p{0.1\textwidth}>{\raggedright}p{0.09\textwidth}>{\raggedright\arraybackslash}p{0.08\textwidth}>{\raggedright\arraybackslash}p{0.15\textwidth}>{\raggedright\arraybackslash}p{0.1\textwidth}}
\toprule
 & Attack Vectors & DAG Structure & Node Attributes & Dynamic Connectors & Edge Attributes & Countermeasure Nodes & Consequence Nodes
\tabularnewline
\midrule
Attack Trees & \checkmark & - & (\checkmark) & - & - & - & -
\tabularnewline
Augmented Vulnerability Trees & \checkmark & - & (\checkmark) & - & - & -  & -
\tabularnewline
Augmented Attack Trees & \checkmark & - & (\checkmark) & - & - & -  & -
\tabularnewline
OWA Trees & \checkmark & - & - & (\checkmark)  & (\checkmark) & -  & -
\tabularnewline
Parallel Model for Multi-Parameter Attack Trees & \checkmark & - & (\checkmark) & - & - & -  & -
\tabularnewline
Extended Fault Trees & \checkmark & - & (\checkmark) & - & - & -  & -
\tabularnewline
\bottomrule
\end{tabular}}
\end{table*}

Each one of the thirty-three frameworks presented in this section considers only subsets of the seven identified properties. 
None of these frameworks are suitable, as all seven properties are necessary to perform a full risk assessment.
To overcome this limitation, this article incorporates all seven identified properties into a framework for a graphical solution for performing risk analysis and examines its applicability to different risk analysis standards.

\subsection{Static Attack Modeling}\label{sec: static attack modeling}

Six frameworks for \emph{static attack modeling}, namely \emph{Attack Trees}~\cite{weiss1991}, \emph{Augmented Vulnerability Trees}~\cite{AugmentedVulnerabilityTrees}, \emph{Augmented Attack Trees}~\cite{AugmentedAttackTrees}, \emph{OWA Trees}~\cite{Yager2006OWATA}, \emph{Parallel Model for Multi-Parameter Attack Trees}~\cite{ParallelModelForMultiParameterAttackTrees}, and \emph{Extended Fault Trees}~\cite{ExtendedFaultTrees}, are summarised in Table~\ref{tab: static attack modeling}.
All frameworks fulfill the attack vectors property, but none of them supports the DAG structure, countermeasure nodes, and consequence nodes properties. 
Of the six frameworks, OWA trees stand out as they at least partially fulfill the dynamic connectors and edge attributes properties, despite being the only framework that does not fulfill the node attributes property. 
This section describes attack trees, which was the first graphical security modeling framework, and OWA trees, which is the framework that at least partially fulfills most of the seven identified properties.

\subsubsection{Attack Trees}\label{sec: attack trees}

The first \emph{tree-based approach}, shown as an AND-OR tree structure for graphical security modeling, was the \emph{threat logic trees}, which was introduced by Weiss in 1991~\cite{weiss1991}.
Today, all AND-OR tree structures are referred to as \emph{attack trees}, a term first introduced by Salter et al. in 1998~\cite{Salter1998}.

In attack trees, the root node (i.e., the tree's root) indicates the attack's main goal. 
The main goal is then conjunctively (AND) or disjunctively (OR) refined into sub-goals until they represent basic actions corresponding to atomic components that can be easily understood and quantified. 
Conjunctive refinements indicate that \emph{all} sub-goals need to be fulfilled in order to achieve the main goal, whereas disjunctive refinements indicate that \emph{at least one} sub-goal needs to be fulfilled for achieving the main goal~\cite{weiss1991}.

\subsubsection{OWA Trees}\label{sec: owa trees}

\emph{Ordered weighted averaging (OWA) trees} were proposed by Yager in 2005 to include the concept of \emph{uncertainty} into attack trees~\cite{Yager2006OWATA}. 
This was made possible by replacing the AND-OR nodes with OWA nodes (i.e., quantifiers, such as \emph{most}, \emph{some}, \emph{half of}, etc.) and therefore taking into consideration situations where the number of sub-goals that need to be fulfilled in order to achieve the main goal remains unknown. 
Finally, OWA trees allow for the evaluation of success probability and cost attributes, which can be jointly used to calculate the cheapest and most probable attack.

\subsection{Sequential Attack Modeling}\label{sec: sequential attack modeling}

\begin{table*}[h]
\rowcolors{2}{gray!10}{gray!40}
\renewcommand{\arraystretch}{1.2}
\caption{Sequential attack modeling frameworks compared to the seven defined properties.}
\label{tab: sequential attack modeling}
\noindent\makebox[\textwidth]{%
\begin{tabular}[t]{>{\raggedright}p{0.15\textwidth}>{\raggedright}p{0.06\textwidth}>{\raggedright}p{0.08\textwidth}>{\raggedright}p{0.1\textwidth}>{\raggedright}p{0.09\textwidth}>{\raggedright\arraybackslash}p{0.08\textwidth}>{\raggedright\arraybackslash}p{0.15\textwidth}>{\raggedright\arraybackslash}p{0.1\textwidth}}
\toprule
 & Attack Vectors & DAG Structure & Node Attributes & Dynamic Connectors & Edge Attributes & Countermeasure Nodes & Consequence Nodes
\tabularnewline
\midrule
Cryptographic DAGs & \checkmark & \checkmark & - & - & - & - & - 
\tabularnewline
Fault Trees for Security & \checkmark & - & \checkmark & (\checkmark) & - & -  & -
\tabularnewline
Bayesian Networks for Security & \checkmark & \checkmark & \checkmark & - & \checkmark & - & -
\tabularnewline
Bayesian Attack Graphs & \checkmark & \checkmark & \checkmark & - & \checkmark & - & -
\tabularnewline
Compromise Graphs & \checkmark & \checkmark & - & - & (\checkmark) & - & -
\tabularnewline
Enhanced Attack Trees & \checkmark & - & \checkmark & - & (\checkmark) & - & -
\tabularnewline
Vulnerability Cause Graphs & (\checkmark) & \checkmark & - & - & - & - & -
\tabularnewline
Dynamic Fault Trees for Security & \checkmark & - & (\checkmark) & - & - & - & -
\tabularnewline
Serial Model for Multi-Parameter Attack Trees & \checkmark & - & (\checkmark) & - & - & - & -
\tabularnewline
Improved Attack Trees & \checkmark & - & (\checkmark) & - & - & - & -
\tabularnewline
Time-dependent Attack Trees & \checkmark & \checkmark & (\checkmark) & - & - & - & -
\tabularnewline
\bottomrule
\end{tabular}}
\end{table*}

Eleven frameworks for \emph{sequential attack modeling}, namely \emph{Cryptographic DAGs}~\cite{Meadows1996ARO}, \emph{Fault Trees for Security}~\cite{FaultTreesForSecurity}, \emph{Bayesian Networks for Security}~\cite{BayesianNetworksForSecurity}, \emph{Bayesian Attack Graphs}~\cite{BayesianAttackGraphs}, \emph{Compromise Graphs}~\cite{CompromiseGraphs}, \emph{Enhanced Attack Trees}~\cite{EnhancedAttackTrees}, \emph{Vulnerability Cause Graphs}~\cite{VulnerabilityCauseGraphs}, \emph{Dynamic Fault Trees for Security}~\cite{DynamicFaultTreesForSecurity}, \emph{Serial Model for Multi-Parameter Attack Trees}~\cite{SerilModelForMultiParameterAttackTrees}, \emph{Improved Attack Trees}~\cite{ImprovedAttackTrees}, and \emph{Time-dependent Attack Trees}~\cite{TimeDependentAttackTrees}, are summarised in Table~\ref{tab: sequential attack modeling}. 
Again, none of the frameworks fulfills the countermeasure and consequence nodes property. In addition, only Fault Trees for Security offer a wide range of dynamic connectors, and only Bayesian-based models fulfill the edge attributes property. 
Finally, Compromise Graphs and Enhanced Attack Trees are two frameworks that at least partially fulfill the edge attributes property, and Vulnerability Cause Graphs are the only framework that only partially fulfills the attack vectors property. 
This section describes Cryptographic DAGs, which was the first graph-based approach for security modeling, and Bayesian Attack Graphs, which combine attack trees and Bayesian networks and also fulfill four of the seven identified properties.

\subsubsection{Cryptographic DAGs}\label{sec: cryptographic dags}

\emph{Cryptographic directed acyclic graphs} were proposed by Meadows~\cite{Meadows1996ARO} in 1996 to provide a \emph{novel} simple representation of sequences and dependencies of attack steps towards the main goal of the attack. 
Instead of a tree-based approach, Cryptographic DAGs introduced a \emph{graph-based approach} for security modeling. 
However, they eventually do not offer the possibility to perform risk assessment as other properties are still not fulfilled.

\subsubsection{Bayesian Networks and Bayesian Attack Graphs}\label{sec: bayesian Networks and Attack Graphs}

For the last couple of decades, researchers have been focusing on \emph{Bayesian networks} for the purposes of security modeling.
The origin of Bayesian networks, which are also known as \emph{belief} or \emph{causal networks}, lies in artificial intelligence.
In Bayesian networks, nodes represent events or objects and are associated with probabilistic variables. 
Hence, analyzing the uncertainty of events is also possible. 
Bayesian networks follow a DAG structure, where the directed edges represent the causal dependencies between the nodes~\cite{BayesianAttackGraphs}.

\emph{Bayesian attack graphs} are a fusion of (general) attack trees and (computational procedures) of Bayesian networks, and they were first introduced by Liu and Man in 2005 to analyze network vulnerability scenarios~\cite{BayesianAttackGraphs}. 
Subsequently, calculating general security metrics regarding information system networks~\cite{Frigault2008, Noel2010} and capturing dynamic behavior~\cite{Frigault2008Dyn} was also made possible.

Finally, although Bayesian attack graphs allow for assigning values to nodes and for performing computations using the graphs, they do not allow for a dynamic selection of connectors and for including countermeasures. 
As a result, Bayesian attack graphs cannot be used to perform risk assessment.

\subsection{Static Attack and Defense Modeling}\label{sec: static attack and defense modeling}

\begin{table*}[h]
\rowcolors{2}{gray!10}{gray!40}
\renewcommand{\arraystretch}{1.2}
\caption{Static attack and defense modeling frameworks compared to the seven defined properties.}
\label{tab: static attack and defense modeling}
\noindent\makebox[\textwidth]{%
\begin{tabular}[t]{>{\raggedright}p{0.15\textwidth}>{\raggedright}p{0.06\textwidth}>{\raggedright}p{0.08\textwidth}>{\raggedright}p{0.1\textwidth}>{\raggedright}p{0.09\textwidth}>{\raggedright\arraybackslash}p{0.08\textwidth}>{\raggedright\arraybackslash}p{0.15\textwidth}>{\raggedright\arraybackslash}p{0.1\textwidth}}
\toprule
 & Attack Vectors & DAG Structure & Node Attributes & Dynamic Connectors & Edge Attributes & Countermeasure Nodes & Consequence Nodes
\tabularnewline
\midrule
Anti-Models & \checkmark & - & - & - & - & \checkmark & -
\tabularnewline
Defense Trees & \checkmark & - & (\checkmark) & - & - & \checkmark & -
\tabularnewline
Protection Trees & - & - & \checkmark & - & - & \checkmark & -
\tabularnewline
Security Activity Graphs & \checkmark & \checkmark & (\checkmark) & - & - & \checkmark & -
\tabularnewline
Attack Countermeasure Trees & \checkmark & - & \checkmark & - & - & \checkmark & -
\tabularnewline
Attack-Defense Trees & \checkmark & - & \checkmark & - & - & \checkmark & -
\tabularnewline
Countermeasure Graphs & \checkmark & \checkmark & \checkmark & - & - & \checkmark & -
\tabularnewline
\bottomrule
\end{tabular}}
\end{table*}

Seven frameworks for \emph{static attack and defense modeling}, namely \emph{Anti-Models}~\cite{AntiModels}, \emph{Defense Trees}~\cite{DefenseTrees}, \emph{Protection Trees}~\cite{ProtectionTrees}, \emph{Security Activity Graphs}~\cite{SecurityActivityGraphs}, \emph{Attack Countermeasure Trees}~\cite{AttackCountermeasureTrees}, \emph{Attack-Defense Trees}~\cite{AttackDefenseTrees}, and \emph{Countermeasure Graphs}~\cite{CountermeasureGraphs}, are summarised in Table~\ref{tab: static attack and defense modeling}. 
All frameworks fulfill the countermeasure nodes property, and only Protection Trees do not fulfill the attack vectors property. 
In addition, Anti-Models is the only framework that does not at least partially fulfill the node attributes property.
However, none of these frameworks considers consequence nodes in their design.
This section describes Security Activity Graphs and Countermeasure Graphs, which are the two frameworks that fulfill four of the seven identified properties.

\subsubsection{Security Activity Graphs}\label{sec: security activity graphs}

\emph{Security activity graphs (SAGs)} were developed by Ardi et al.~\cite{SecurityActivityGraphs} in 2006 to improve security throughout the software development process. 
SAGs are loosely based on fault trees, and the root of a SAG is associated with a vulnerability. 
Vulnerability mitigations are modeled using activities (i.e., leaf nodes), which are assigned boolean variables to indicate whether an activity \enquote{is implemented perfectly during software development} (true) or not (false). 
Finally, besides AND-OR gates, which follow a strictly Boolean logic, SAGs also include \emph{split gates}, which allow one activity to be used in several parent activities, thus creating a DAG structure.

However, SAGs lack the ability to represent the consequences of threats and edge attributes, and both are necessary to calculate a risk value.
Furthermore, there is only a limited option for connectors and node attributes.
Therefore, rendering SAGs impractical for risk assessment.

\subsubsection{Countermeasure Graphs}\label{sec: countermeasure graphs}

\emph{Countermeasure graphs} were introduced by Baca and Petersen~\cite{CountermeasureGraphs} in 2010 to simplify countermeasure selection through cumulative voting. 
Countermeasure graphs are created by identifying actors, goals, attacks, and countermeasures. Related events are connected with edges. 
That is, actors are connected to pursued goals and likely executable attacks, and countermeasures are connected to preventable attacks. 
Finally, actors, goals, attacks, and countermeasures are assigned priorities according to the rules of hierarchical cumulative voting. 
Higher assigned priorities imply higher threat levels of the corresponding events and vice versa. 
Using hierarchical cumulative voting, the most effective countermeasures can be identified.

Countermeasure Graphs provide a useful system overview, but the computational rules focus on finding the most effective countermeasure instead of the most likely and severe attack. 
This limitation could at least partially be addressed with the threat level. 
However, the threat level value is determined by the subjective assessment of the graph creator rather than by calculations over meaningful attributes, thereby raising issues of validity.

\subsection{Sequential Attack and Defense Modeling}\label{sec: sequential attack and defense modeling}

\begin{table*}[h]
\rowcolors{2}{gray!10}{gray!40}
\renewcommand{\arraystretch}{1.2}
\caption{Sequential attack and defense modeling frameworks compared to the seven defined properties.}
\label{tab: sequential attack and defense modeling}
\noindent\makebox[\textwidth]{%
\begin{tabular}[t]{>{\raggedright}p{0.15\textwidth}>{\raggedright}p{0.06\textwidth}>{\raggedright}p{0.08\textwidth}>{\raggedright}p{0.1\textwidth}>{\raggedright}p{0.09\textwidth}>{\raggedright\arraybackslash}p{0.08\textwidth}>{\raggedright\arraybackslash}p{0.15\textwidth}>{\raggedright\arraybackslash}p{0.1\textwidth}}
\toprule
 & Attack Vectors & DAG Structure & Node Attributes & Dynamic Connectors & Edge Attributes & Countermeasure Nodes & Consequence Nodes
\tabularnewline
\midrule
Insecurity Flows & \checkmark & \checkmark & \checkmark & - & - & \checkmark & -
\tabularnewline
Intrusion DAGs & \checkmark & \checkmark & - & - & - & \checkmark & -
\tabularnewline
Bayesian Defense Graphs & \checkmark & \checkmark & (\checkmark) & - & - & \checkmark & -
\tabularnewline
Security Goal Indicator Trees & - & - & - & - & - & \checkmark & -
\tabularnewline
Attack Response Trees & \checkmark & -  & \checkmark & - & - & \checkmark & -
\tabularnewline
Boolean Logic Driven Markov Processes & \checkmark & \checkmark & (\checkmark) & (\checkmark) & - & \checkmark & -
\tabularnewline
Cyber Security Modeling Language & \checkmark & \checkmark & (\checkmark) & - & (\checkmark) & \checkmark & -
\tabularnewline
Security Goal Models & \checkmark & \checkmark & - & - & - & \checkmark & -
\tabularnewline
Unified Parameterizable Attack Trees & \checkmark & - & \checkmark & - & (\checkmark) & \checkmark & -
\tabularnewline
\bottomrule
\end{tabular}}
\end{table*}

Finally, nine frameworks for \emph{sequential attack and defense modeling}, namely \emph{Insecurity Flows}~\cite{InsecurityFlows}, \emph{Intrusion DAGs}~\cite{IntrusionDAGs}, \emph{Bayesian Defense Graphs}~\cite{BayesianDefenseGraphs}, \emph{Security Goal Indicator Trees}~\cite{SecurityGoalIndicatorTrees}, \emph{Attack Response Trees}~\cite{AttackResponseTrees}, \emph{Boolean Logic Driven Markov Process}~\cite{BooleanLogicDrivenMarkovProcess}, \emph{Cyber Security Modeling Language}~\cite{CyberSecurityModelingLanguage2010}, \emph{Security Goal Models}~\cite{SecurityGoalModels}, and \emph{Unified Parameterizable Attack Trees}~\cite{UnifiedParameterizableAttackTrees}, are summarized in Table~\ref{tab: sequential attack and defense modeling}. 
All frameworks fulfill the countermeasure nodes property, and only Security Goal Indicator Trees do not fulfill the attack vectors property. 
In addition, Boolean Logic Driven Markov Processes (BDMPs) is the only framework that offers a wide range of connectors and, therefore, at least partially fulfills the dynamic connectors property. 
This section describes BDMPs and Cyber Security Modeling Language (CySeMoL), which are the two frameworks that at least partially fulfill five of the seven identified properties.

\subsubsection{Boolean Logic Driven Markov Processes}\label{sec: boolean logic driven markov processes}

\emph{Boolean logic driven Markov processes (BDMPs)} are a security modeling framework, which can also be used to perform risk assessment~\cite{BooleanLogicDrivenMarkovProcess}. 
It was invented by Bouissou and Bon~\cite{BooleanLogicDrivenMarkovProcess} in 2003 for the safety and reliability domains, and it was later adapted to security modeling by Piètre-Cambacédès and Bouissou in 2010. 
BDMPs combine the readability of attack trees with the modeling power of Markov chains. 
The root (top event) of a BDMP represents the main goal of the attack, and the leaves represent the attack steps or security events.
BDMPs offer a wide range of node attributes, including time-domain metrics, such as mean-time to success, attack tree-related metrics, such as costs of attacks, boolean indicators, such as specific requirements, and risk assessment tools, such as sensibility graphs.

However, the lack of edge attributes, in addition to issues of usability with respect to leaf nodes and connectors~\cite{BDMPCritic}, render BDMPs impractical for risk assessment.

\subsubsection{Cyber Security Modeling Language}\label{sec: cyber security modeling language}

\emph{Cyber security modeling language (CySeMoL)} was developed by Sommestad et al. in 2010 to assess the cyber security of \emph{supervisory control and data acquisition (SCADA)} system architectures~\cite{CyberSecurityModelingLanguage2010, CyberSecurityModelingLanguage2013}.
Simply modeling the system architecture and the characteristics of the involved assets is sufficient, as CySeMoL already includes information about how attacks and defenses are quantitatively related. 
The attacker is assumed to be a professional penetration tester with a fixed time of one week to perform an attack.
CySeMoL was extended by Holm in 2014 and renamed to \emph{predictive, probabilistic cyber security modeling language ((P$^2$)CySeMoL)}, introducing more flexible and useful computations, the possibility to model assets, attacks, and defenses that are not necessarily SCADA-related, and the option to specify the time needed to perform an attack~\cite{PredictiveProbabilisticCyberSecurityModelingLanguage}.
Computations can be conducted automatically (i.e., without personalized inputs) as (P$^2$)CySeMoL already includes qualitative information gathered from literature reviews, empirical studies, as well as surveys involving domain experts~\cite{CyberSecurityModelingLanguage2010, CyberSecurityModelingLanguage2013, PredictiveProbabilisticCyberSecurityModelingLanguage}.

The results of the computations show the likelihood of an attack. 
However, the severity of an attack is not considered, and therefore the risk of an attack cannot be properly assessed. 
Furthermore, (P$^2$)CySeMoL does not include connectors, and therefore it seems an inconvenient tool for graphical risk assessment.

\subsection{Summary of Remarks}\label{sec2: summary of remarks}

This section provides an overview of thirty-three frameworks for analysis of attack and defense scenarios, and it describes eight of these frameworks in more detail. 
Thirty frameworks fulfill the attack vectors property, sixteen frameworks fulfill the countermeasure nodes property, and only thirteen frameworks fulfill the DAG structure property.
In addition, node/edge attributes and connectors are, in most cases, fixed and limited, thereby reducing the usability and usefulness of the frameworks with respect to the purposes of risk assessment. 
The complex nature and rapid development of (information) systems, attacks, and defenses motivate the need for proper risk management.
Existing methods are mainly consisting of tables with graphical solutions mostly utilized for support, if at all.
As shown in this section, current graphical solutions support threat or vulnerability management and sometimes even calculations to determine which attack vector might be the easiest to execute or, in other terms, which is most probable to occur.
The risk value cannot be equated with probability, though, and is usually determined using the probability of an event and its impact.
However, none of the methods described in this section can represent an event's consequences and impact, rendering them incapable of performing risk assessment.


\section{Attack Graphs}\label{sec: attack graphs}

\begin{figure*}[ht]
\includegraphics[width=\textwidth]{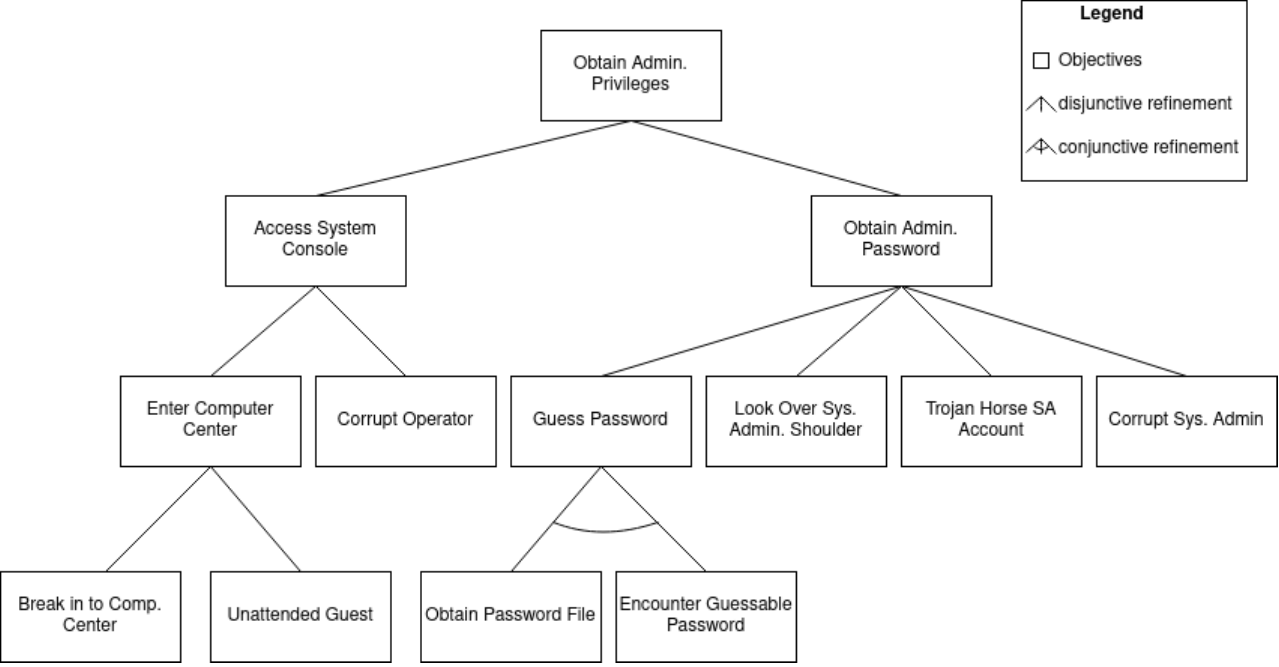}
\caption{Attack tree about obtaining administrator privileges on a UNIX system, designed by Weiss in 1991~\cite{weiss1991}.}
\label{fig: attack tree weiss}
\end{figure*}

This article proposes the use of a \emph{DAG structure} for the development of the security modeling framework. 
DAGs consist of nodes, which are connected through directed edges that do not form any loops. 
The main DAG components are illustrated in Weiss's attack tree~\cite{weiss1991} shown in Figure~\ref{fig: attack tree weiss}.

\subsection{Attack Graph Components}\label{sec: attack graph components}

This section summarises the main components of an Attack Graph.\\

\noindent
\textbf{Nodes}\\
\emph{Nodes} represent the attacker's goals. 
There are three different types of nodes: (1) Root Nodes, the set of which is defined as $\mathcal{R}$, (2) Leaf Nodes, the set of which is defined as $\mathcal{L}$, and (3) Inner Nodes, the set of which is defined as $\mathcal{I}$.
Finally, all nodes $\emph{n} \in \mathcal{N}$ form the set $\mathcal{N}$:

\begin{center}
$\mathcal{N} = \mathcal{R} \cup \mathcal{L} \cup \mathcal{I}$.
\end{center}

\noindent
\textbf{Edges}\\
\emph{Edges} indicate relations between nodes. 
Attack Graphs use directed edges, thereby enabling nodes to have \emph{predecessors} (also called \emph{parents} or \emph{ancestors}) and/or \emph{successors} (also called \emph{children}). 
The set of predecessors of a node $n \in \mathcal{N}$ is defined as $Pred_n$, and the set of successors of a node $n \in \mathcal{N}$ is defined as $Succ_n$. 
For example, for node $n$ := \enquote{Access System Console} in Figure~\ref{fig: attack tree weiss}, $Pred_n$ = \{\enquote{Obtain Admin. Privileges}\} and $Succ_n$ = \{\enquote{Enter Computer Center}, \enquote{Corrupt Operator}\}.\\

\noindent
\textbf{Root Nodes}\\
A node that has no predecessor is a Root Node $\emph{r} \in \mathcal{R}$. \emph{Root Nodes} indicate the main goal of the attack. 
For example, the Root Node in Figure~\ref{fig: attack tree weiss} is \enquote{Obtain Admin. Privileges}.\\

\noindent
\textbf{Leaf Nodes}\\
A node that has no successor is a Leaf Node $\emph{l} \in \mathcal{L}$. 
\emph{Leaf Nodes} indicate basic actions that can easily be understood and quantified. 
For example, three Leaf Nodes in Figure~\ref{fig: attack tree weiss} are \enquote{Break in to Comp. Center}, \enquote{Obtain Password File}, and \enquote{Corrupt Sys. Admin}.\\

\noindent
\textbf{Inner Nodes}\\
A node that has at least one predecessor and at least one successor is an Inner Node $\emph{i} \in \mathcal{I}$. 
On one hand, \emph{Inner Nodes} are specialized actions, goals, or sub-goals of their predecessor nodes. 
On the other hand, they are generalized actions or goals of their successor nodes. 
For example, two Inner Nodes in Figure~\ref{fig: attack tree weiss} are \enquote{Access System Console} and \enquote{Enter Computer Center}.\\

\noindent
\textbf{Node Attributes}\\
Nodes are refined into sub-goals until they represent basic actions. 
Basic actions can then be quantified by assigning \emph{attributes}, which are quantifiable properties, to nodes.
A set of node attributes is defined as $\mathcal{A}$. 
For every $\emph{a}_k \in \mathcal{A}$, there exists a value $\emph{v}_m \in \mathcal{V}_k $, where $\mathcal{V}_k$ is the finite set of acceptable values of the attribute $\emph{a}_k \in \mathcal{A}$. 
The process of assigning values to an attribute is defined by the function $g$:

\begin{center}
$g(\emph{a}_k) = \emph{v}_m$,
\end{center}

\noindent
where $\emph{a}_k \in \mathcal{A}$ is an attribute of a node $\emph{n} \in \mathcal{N} $, and $\emph{v}_m \in \mathcal{V}_k$ is a value of this attribute. 
The set $\mathcal{F}_g$ contains all functions $g$ that assign values to attributes.\\


\noindent
\textbf{Edge Attributes}\\
The relation between nodes can be quantified by assigning attributes to edges. 
A set of edge attributes is defined as $\mathcal{A}_\mathcal{E}$. Edge attributes cannot be assigned to nodes, and similarly, node attributes cannot be assigned to edges.
Therefore, $\mathcal{A}_\mathcal{E} \cap \mathcal{A} = \emptyset$.\\

\noindent
\textbf{Connectors}\\
\emph{Connectors} $\emph{c} \in \mathcal{C}$ indicate the refinements of nodes. 
Conjunctive refinements (AND) indicate that \emph{all} successor nodes need to be fulfilled to achieve the refined node's goal.
Disjunctive refinements (OR) indicate that \emph{at least one} successor node needs to be fulfilled to achieve the goal of the refined node.\\


\noindent
\textbf{Aggregated Attributes}\\
By definition, a node $n \in \mathcal{N}$ and all successor nodes $Succ_n$ have the same set of attributes. 
Hence, the attribute values of node $n \in \mathcal{N}$ are determined by aggregating the attribute values of all of its successor nodes $Succ_n$. 

For example, let $n_2$ and $n_3$ be successor nodes of $n_1$, with $n_1$, $n_2$, $n_3 \in \mathcal{N}$.
Further, let $a_1$ be an attribute from the set $\mathcal{A}$.
Then, there is a set $\mathcal{V}_1$ with acceptable values for the attribute $a_1$ and all three nodes have this attribute
To determine the value for $a_1$ of node $n_1$, the values of $a_1$ of the nodes $n_2$ and $n_3$ are aggregated using a function $f$.
Whereas the result value needs to be in the set of acceptable values $\mathcal{V}_1$ for $a_1$.

An attribute value can be typically computed using a function that returns (1) the \emph{maximum}, (2) the \emph{minimum}, (3) the \emph{sum}, or (4) the \emph{product} of a set of given attribute values. 
In this case, the set of attribute values $v_m \in \mathcal{V}_k$ needs to be arranged in a natural sort order, where $x_{min}$ is the smallest and $x_{max}$ is the largest value of the set. 
The following functions are commonly utilized for aggregating attributes. 
However, our framework is open to other functions as well.
In the following definitions, $a_1$ represents the attribute for node $n_1$, $a_2$ represents the attribute for node $n_2$, and so on.
However, all attributes $a_1$, $a_2, ...$ represent the same property, e.g., knowledge needed to perform this action.

\smallskip
\noindent
\textbf{Maximum.} The maximum function takes only the highest value:
\begin{center}
$f_{max}(g(a_1), g(a_{2}), ..., g(a_{k})) = \max\{ g(a_1), g(a_{2}), ..., g(a_{k})\}$.
\end{center}

\noindent
\textbf{Minimum.} The minimum function takes only the lowest value:
\begin{center}
$f_{min}(g(a_1), g(a_{2}), ..., g(a_{k})) = \min\{ g(a_1), g(a_{2}), ..., g(a_{k})\}$.
\end{center}

\noindent
\textbf{Sum.} The sum function adds the values:
\begin{center}
$f_{sum}(g(a_1), g(a_{2}), ..., g(a_{k})) = \min\{\sum\limits_{j = 1}^{k} g(a_j), x_{max}\}$.
\end{center}

\noindent
\textbf{Product.} The product function multiplies the values:
\begin{center}
$f_{prod}(g(a_1), g(a_{2}), ..., g(a_{k})) = \min\{\prod\limits_{j = 1}^{k} g(a_j), x_{max}\}$.
\end{center}

\subsection{Attack Graph Definition}\label{sec: attack graph definition}

We thus arrive at the following definition of an Attack Graph.\\

\noindent
\emph{An Attack Graph $\mathcal{G}$ is a directed acyclic graph (DAG) containing logical connectors $\mathcal{C}$ and depicting an attack scenario. 
The edges $\mathcal{E}$ are weighted, and the nodes $\mathcal{N}$ include attributes $\mathcal{A}$ that represent the difficulty of performing this attack step. Attributes have a predefined set of values $\mathcal{V}$. 
Finally, functions in the set $F_g$ assign the values to the attributes, and functions in the set $F_f$ aggregate values of successor nodes.}\\

\begin{center}
$\mathcal{G} = \{\mathcal{N},\mathcal{E}, \mathcal{C},\mathcal{A},\mathcal{V},\mathcal{F}_f,\mathcal{F}_g\}$
\end{center}

\section{Attack Graph Risk Assessment}\label{sec: attack graph risk assessment}

\begin{figure*}[ht]
\includegraphics[width=\textwidth]{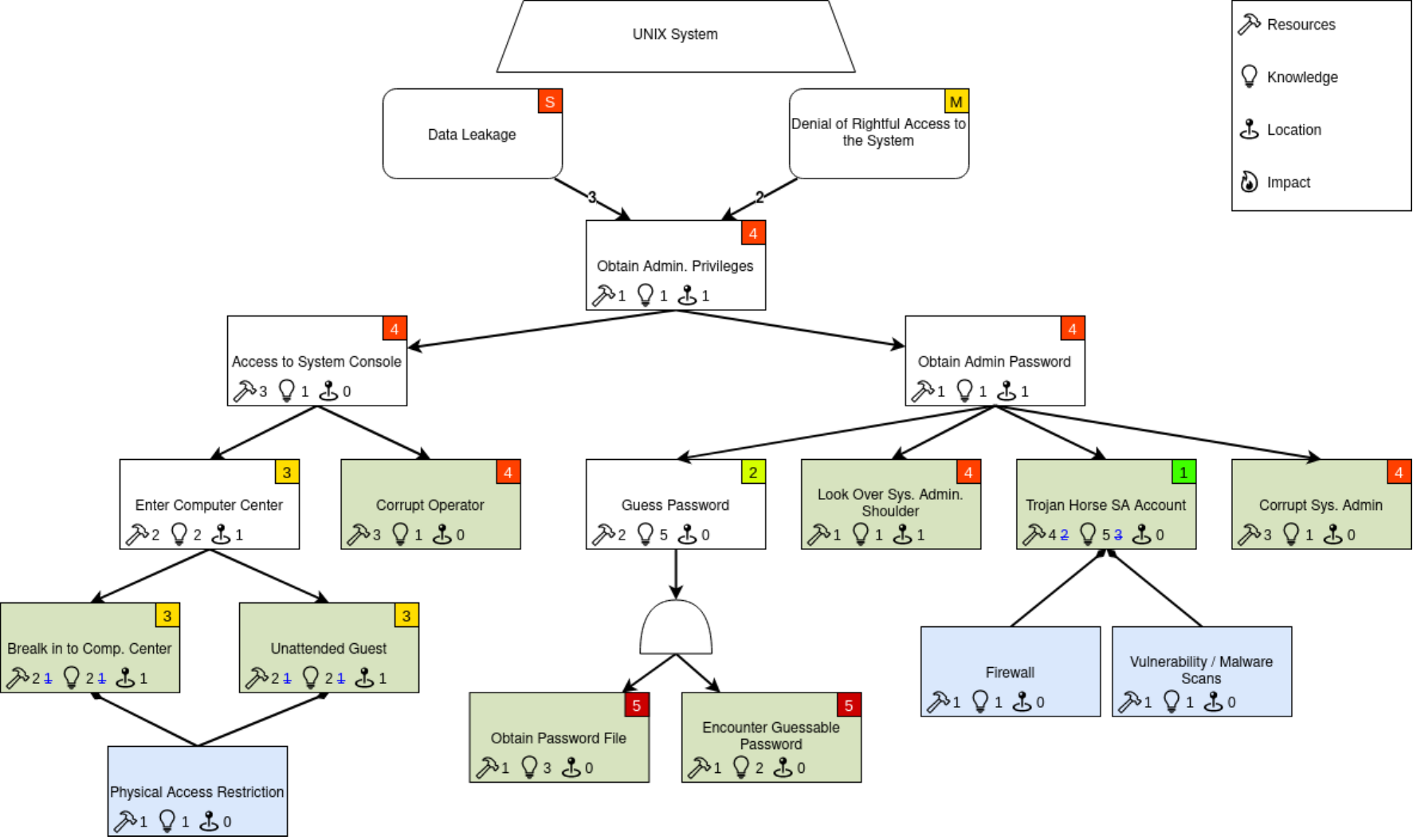}
\caption{Attack tree about obtaining administrator privileges on a UNIX system, enhanced with the six properties proposed in this article.}
\label{fig: attack tree weiss modified}
\end{figure*}

We enhance the Attack Graph components summarized in Section~\ref{sec: attack graphs} with (1) \emph{Consequence Nodes}, which indicate the negative consequences of the main goal of the attack and create a relation to constitute the impact of an attack, (2) \emph{Attack Feasibility Attributes}, which indicate the ease of launching an attack, and (3) \emph{Countermeasure Nodes}, which indicate security measures for either preventing or mitigating attack steps or the impact of those.
The Attack Graph adaptation of Weiss's attack tree is shown in Figure~\ref{fig: attack tree weiss modified}.

\noindent
\textbf{Consequence Nodes}\\
In Attack Graphs, \emph{Consequence Nodes} (i.e., \enquote{Data Leakage} and \enquote{Denial of Rightful Access to the System}), shown as rounded corner rectangles, become the highest nodes (i.e. the Root Nodes) in the tree structure.
They indicate the negative consequences of the main goal of the attack, which is now represented by the topmost Inner Node (i.e., \enquote{Obtain Admin. Privileges}). 
For the rest of the article, Root Nodes are referred to as Consequence Nodes.

Consequence nodes are not part of the set $\mathcal{N}$ and build their own set $\mathcal{N}_C$, with $\mathcal{N} \cap \mathcal{N}_C = \emptyset$, and with a different set of attributes $\mathcal{A}_R$, with $\mathcal{A}_R \cap \mathcal{A} = \emptyset$.
This is necessary as nodes of the set $\mathcal{N}$ represent actions with their attributes, consequently describing properties necessary to perform the action or how easy it is to perform this action.
Whereas nodes of the set $\mathcal{N}_C$ describe the consequences of actions, and thus their attributes  typically describe the risk of that consequence to arise. \\

\noindent
\textbf{Attack Feasibility Attributes}\\ 
\emph{Attack Feasibility Attributes} are shown as colored squares on the top right of Inner Nodes, indicating the ease of launching an attack.
They are computed using either (1) a \emph{function} of a set of given attribute values, (2) a \emph{matrix} of a set of given attribute values or (3) a combination of these methods.

In the first case, a value $\emph{v}_m \in \mathcal{V}_k$ of an Attack Feasibility Attribute $\emph{a}_k \in \mathcal{A}$ of a node $n$ can be computed using a function that aggregates the attributes $\emph{a}_{1}, \emph{a}_{2}, ..., a_{k-1} \in \mathcal{A}$ of node $n$, with $a_k \neq a_{1}, a_{2}, ..., a_{k-1}$:

\begin{center}
$\emph{v}_m = f(\emph{a}_{1}, \emph{a}_{2}, ..., a_{k-1})$.
\end{center}

\noindent
Similarly, in the second case, a set of attributes $d := |\{a_1, a_{2}, ..., a_{k-1}\}|$ are aggregated over a $d$-dimensional matrix. 
An example of a 2-dimensional matrix is shown in Table~\ref{tab: 2d matrix}. 
Here, for a node $n$ with attributes $a_1$ and $a_2$, there exist a set of values $\mathcal{V}_1 := \{v_0, v_1, ..., v_{m_1}\}$ for $a_1$ and a set of values $\mathcal{V}_2 := \{u_0, u_1, ..., u_{m_2}\}$ for $a_2$. 
As such, the aggregation of $\mathcal{V}_1$ and $\mathcal{V}_2$ returns a set of values $\mathcal{V}_3 := \{v_{00}, v_{01}, ..., v_{{m_1}{m_2}}\}$ for an Attack Feasibility Attribute $a_3$.

\begin{table}[h]
\renewcommand{\arraystretch}{1.2}
\caption{2D matrix for an Attack Feasibility Attribute $a_3$, with $a_1, a_2, a_3$ being attributes of a node $n$.}
\label{tab: 2d matrix}
\noindent\makebox[\linewidth]{%
\begin{tabular}[t]{>{\raggedright}p{0.05\textwidth}>{\raggedright}p{0.05\textwidth}>{\raggedright}p{0.05\textwidth}>{\raggedright}p{0.05\textwidth}>{\raggedright\arraybackslash}p{0.05\textwidth}}
\toprule
 & \multicolumn{4}{c}{$a_2$}
\tabularnewline
\cmidrule{2-5}
$a_1$ & $u_0$ & $u_1$ & ... & $u_{m_2}$
\tabularnewline
\midrule
$v_0$ & $v_{00}$ & $v_{01}$ & ... & $v_{0{m_2}}$
\tabularnewline
$v_1$ & $v_{10}$ & $v_{11}$  & ... & $v_{1{m_2}}$
\tabularnewline
... & ... & ... & ... & ...
\tabularnewline
$v_{m_1}$ & $v_{{m_1}0}$ & $v_{{m_1}1}$ & ... & $v_{{m_1}{m_2}}$
\tabularnewline
\bottomrule
\end{tabular}}
\end{table}

In the third case, the two previously described methods are combined in an arbitrary way.
An example that is also used later, is that the (genuine) subset $s \subset \{a_1, a_{2}, ..., a_{k-1}\}$ is aggregated using an $|s|$-dimensional matrix, which generates a preliminary result $v_{pre}$.
The remaining attributes $r := \{a_1, a_{2}, ..., a_{k-1}\} \setminus s$ are then aggregated together with $v_{pre}$ in a function $f$, generating the final result.
E.g., let s be the set of values $\{a_1, a_{2}, ..., a_{k-2}\}$ which are aggregated to form the preliminary result $v_{pre}$.
Then $r = \{a_{k-1}\}$ and the result $v$ is calculated by using the function $f$ over the preliminary result $v_{pre}$ and the value of $a_{k-1}$: $v = f(v_{pre}, g(a_{k-1}))$.

We want to acknowledge the fact that the \emph{Attack Feasibility} attribute can have different names, like attack or success probability or likelihood, especially in some standards.
However, \emph{probability} is not a useful name and can lead to some irritation, as it is not mathematically computable probability in terms of statistics or percentages.
Hence, we chose a different wording that is also present in some standards.\\


\noindent
\textbf{Countermeasure Nodes}\\
Finally, \emph{Countermeasure Nodes} (i.e., \enquote{Physical Access Restriction}, \enquote{Firewall} and \enquote{Vulnerability / Malware Scans}), shown as light blue nodes in~\cref{fig: attack tree weiss modified}, are included to indicate the security measures taken to mitigate the risk.
There are two ways how those countermeasures can influence the risk. 
First, there are countermeasures that mitigate or prevent specific attack steps. 
These are attached as new leaf nodes, as shown in~\cref{fig: attack tree weiss modified}. 
Using the aggregation functions, the node attributes of the parent nodes are influenced. 
In the example of~\cref{fig: attack tree weiss modified} the values of the countermeasure nodes are added to their parent nodes which complicate the attack and therefore lower the attack feasibility of the predecessor nodes. 
Taking the countermeasure node \enquote{Physical Access Restriction} and the lowermost inner node \enquote{Break in to Comp. Center} on the lower left of~\cref{fig: attack tree weiss modified}.
The node \enquote{Physical Access Restriction} is rated to increase the value of \emph{Resources} and \emph{Knowledge} by one as the aggregation function chosen here is the sum function.
Hence, the values for \emph{Resources} and \emph{Knowledge} of \enquote{Break in to Comp. Center} are both increased by one to their new value of two.
The originally rated value for \enquote{Break in to Comp. Center} is still visible in blue but crossed out in~\cref{fig: attack tree weiss modified} to illustrate the effect of countermeasure nodes.

Second, the countermeasure can influence an attack's impact on a specific consequence. 
This case is a bit more complicated to illustrate as it is influencing an edge (between a \emph{Consequence Node} and a topmost Inner Node) instead of a node. 
However, the principle is the same as for nodes.

\subsection{Application of Risk Assessment Graphs}\label{sec: attack graph risk estimation}

We now show an example of how to adapt the method of Attack Graphs to facilitate risk management.
When the described definitions are used to create a graphical representation or risk landscape, we refer to these graphs as \emph{Risk Assessment Graphs}.

\begin{table*}[ht]
\renewcommand{\arraystretch}{1.2}
\caption{Resources, Knowledge, and Location matrix.}
\label{tab: resources, knowledge, and location matrix}
\noindent\makebox[\textwidth]{%
\begin{tabular}[t]{>{\raggedright}p{0.15\textwidth}>{\centering}p{0.1\textwidth}>{\centering}p{0.15\textwidth}>{\centering}p{0.1\textwidth}>{\centering}p{0.1\textwidth}>{\centering}p{0.1\textwidth}>{\centering\arraybackslash}p{0.15\textwidth}}
\toprule
 & 0 & 1 & 2 & 3 & 4 & 5
\tabularnewline
\midrule
Resources & - & Basic & Low & Medium & High & Extraordinary
\tabularnewline
Knowledge & - & Basic & Low & Medium & High & Extraordinary
\tabularnewline
Location & Remote & Local & - & - & - & -
\tabularnewline
\bottomrule
\end{tabular}}
\end{table*}

According to DIN VDE V 0831-104, the \emph{Attack Feasibility (AF)} of the Inner Nodes shown in Figure~\ref{fig: attack tree weiss modified} is computed based on the \emph{attacker's capabilities} and \emph{mitigation factors}. 
First, the attacker's capabilities are described by two attributes: (1) \emph{Resources (R)}, reflecting the financial and workforce capacity of the attacker to prepare and launch an attack, and (2) \emph{Knowledge (K)}, reflecting the information that the attacker holds about the system they intend to attack. 
The Resources and Knowledge of the attacker are each rated as \emph{low} ($R,K = 2$), \emph{medium} ($R,K = 3$), or \emph{high} ($R,K = 4$) in the example given in DIN VDE 0831-104. As such, attackers with \emph{basic} and \emph{extraordinary }capabilities ($R,K = 1$ and $R,K = 5$) are not considered.
The values for Resources and Knowledge were extended to distinguish better between different attacks.
Second, the mitigation factors relate to the risk of the attacker being discovered, and they are described by the \emph{Location (L)} attribute, which reflects whether an attack can be launched remotely ($L = 0$) or locally ($L = 1$). 
The values of the three attributes are summarised in Table~\ref{tab: resources, knowledge, and location matrix}.

In this regard, every lowermost Inner Node (green node) is assigned the Resources (hammer) and Knowledge (light bulb) attributes, whose value range from 2 to 4, and a Location (pin) attribute, whose value is either 0 or 1.
First, the Resources and Knowledge of the attacker are related in a matrix, such as the one shown in Table~\ref{tab: preliminary attack feasibility matrix}, to initially determine a \emph{Preliminary Attack Feasibility (PAF)}, which indicates the ease of launching an attack without taking into consideration the risk of being discovered. 
For example, Figure~\ref{fig: attack tree weiss modified} shows that the \enquote{Break in to Comp. Center} attack (i.e., bottom left Leaf Node) requires \emph{Low Resources} ($R = 2$) and \emph{Low Knowledge} ($K = 2$) to be launched. 
In this case, Table~\ref{tab: preliminary attack feasibility matrix} shows that for $R = 3$ and $K = 3$, $PAF = 4$. 
Second, the Location of the attacker is subtracted from the Preliminary Attack Feasibility, to eventually determine the Attack Feasibility. 
Here, Figure~\ref{fig: attack tree weiss modified} shows that the \enquote{Break in to Comp. Center} attack requires \emph{Local Access} ($L = 1$) to be launched.
In this case, the Attack Feasibility is equal to 3 ($AF = PAF - L = 4 - 1 = 3$). 

\begin{table}[h]
\renewcommand{\arraystretch}{1.2}
\caption{Preliminary Attack Feasibility matrix.}
\label{tab: preliminary attack feasibility matrix}
\noindent\makebox[\linewidth]{%
\begin{tabular}[t]{>{\raggedright}p{0.05\textwidth}>{\raggedright}p{0.05\textwidth}>{\raggedright}p{0.05\textwidth}>{\raggedright}p{0.05\textwidth}>{\raggedright}p{0.05\textwidth}>{\raggedright\arraybackslash}p{0.025\textwidth}}
\toprule
 & \multicolumn{5}{c}{K}
\tabularnewline
\cmidrule{2-6}
R & 1 & 2 & 3 & 4 & 5
\tabularnewline
\midrule
1 & 5 & 5 & 4 & 3 & 2
\tabularnewline
2 & 5 & 4 & 4 & 3 & 2
\tabularnewline
3 & 5 & 4 & 4 & 3 & 2
\tabularnewline
4 & 4 & 3 & 3 & 2 & 1
\tabularnewline
5 & 3 & 2 & 2 & 1 & 1
\tabularnewline
\bottomrule
\end{tabular}}
\end{table}

When the Attack Feasibility of all lowermost Inner Nodes is determined, the Resources, Knowledge, and Location attributes of all predecessor Inner Nodes are obtained from their successor Inner Nodes, in this case, either conjunctively (AND) or disjunctively (OR).
Other graphs can include different connectors as well.
Similar to Weiss's attack tree~\cite{weiss1991}, only conjunctive refinements are shown explicitly. 
That is, all refinements shown in the Risk Assessment Graph are disjunctive unless a predecessor Inner Node is connected to its successor Inner Node through an AND gate, in which case the refinement is conjunctive. 
Regarding conjunctive refinements, an Inner Node obtains, in this case, the sum of values for each Resource, Knowledge, and Location attribute exhibited among its successor nodes, and the Attack Feasibility is determined based on the resulting values. 
For example, the attack \emph{Guess Password} can only succeed if an attacker obtains the password file and encounters a guessable password.
Hence, the nodes \emph{Obtain Password File} ($R=1, K=3, L=0$) and \emph{Encounter Guessable Password} ($R=1, K=2, L=0$) are conjunctively joined and their predecessor node \emph{Guess Password} obtains the sum of their values ($R=2, K=5, L=0$).
This particular sum function has an upper limit, meaning that if the addition of the successor values would result in a higher value than defined in the range of this attribute, then the maximum value for this attribute will be chosen for the predecessor node.
For example, if the sum of the values for Knowledge would exceed the value 5, the predecessor node will obtain 5 as the value for Knowledge.
Regarding disjunctive refinements, an Inner Node obtains its successor node's Resources, Knowledge, and Location attributes with the highest Attack Feasibility. 
If the highest Attack Feasibility is exhibited by multiple successor nodes, we defined an order on the attributes, deeming Resources more critical than knowledge.
If the lowest Resource value of the successor nodes exhibiting the highest Attack Feasibility is exhibited by multiple successor nodes, the values for Knowledge are compared.
If multiple of these nodes exhibit the same lowest value for Knowledge, any node can be chosen as they must have the same rating.
Because, if the values for Attack Feasibility, Resources, and Knowledge are identical, then the values of Location have to be identical as well or it would not result in the same Attack Feasibility.
For example, in the case of \emph{Look over Sys. Admin. Shoulder} and \emph{Corrupt Sys. Admin}, both have an Attack Feasibility of 4, so to decide which one is more critical, the algorithm checks the Resource Attribute.
The Resource Attribute for those nodes is different and as \emph{Look Over Sys. Admin. Shoulder} requires only \emph{Basic Resources} the predecessor node \emph{Obtain Admin Password} obtains its values. 

\begin{table*}[h]
\renewcommand{\arraystretch}{1.2}
\caption{Impact matrix.}
\label{tab: impact matrix}
\noindent\makebox[\textwidth]{%
\begin{tabular}[t]{>{\raggedright}p{0.05\textwidth}>{\raggedright}p{0.1\textwidth}>{\raggedright\arraybackslash}p{0.7\textwidth}}
\toprule
Value & Impact & Description
\tabularnewline
\midrule
1 & Negligible & Impact can be readily absorbed, without requiring management effort.
\tabularnewline
2 & Minor & Impact can be readily absorbed, requiring some management effort.
\tabularnewline
3 & Moderate & Impact cannot be readily absorbed, requiring a modest level of resources and management effort.
\tabularnewline
4 & Major & Impact requires a high level of resources and management effort to rectify.
\tabularnewline
5 & Severe & Disaster with the potential to lead to business collapse, requiring total management effort to rectify.
\tabularnewline
\bottomrule
\end{tabular}}
\end{table*}

\begin{table*}[h]
\renewcommand{\arraystretch}{1.2}
\caption{Risk matrix.}
\label{tab: risk matrix}
\noindent\makebox[\textwidth]{%
\begin{tabular}[t]{>{\raggedright}p{0.1\textwidth}>{\raggedright}p{0.15\textwidth}>{\raggedright}p{0.15\textwidth}>{\raggedright}p{0.15\textwidth}>{\raggedright}p{0.15\textwidth}>{\raggedright\arraybackslash}p{0.15\textwidth}}
\toprule
\multirow{2}[3]{*}{Impact} & \multicolumn{5}{c}{Attack Feasibility}
\tabularnewline
\cmidrule{2-6}
 & 1 & 2 & 3 & 4 & 5
\tabularnewline
\midrule
1 & Low & Low & Low & Low & Low
\tabularnewline
2 & Low & Low & Moderate & Moderate & Moderate
\tabularnewline
3 & Low & Moderate & Moderate & Significant & Significant
\tabularnewline
4 & Low & Moderate & Significant & Very High & Very High
\tabularnewline
5 & Low & Moderate & Significant & Very High & Very High
\tabularnewline
\bottomrule
\end{tabular}}
\end{table*}

In addition, the edges relating Consequence Nodes to the topmost Inner Node are assigned with an \emph{Impact} attribute, which indicates the magnitude of damage or physical harm caused by negative consequences on the system. 
The impact of negative consequences is typically rated using numerical ranges and qualitative scales, as shown in Table~\ref{tab: impact matrix}, however, the impact rating terminology remains rather inconsistent across risk assessment standards. 
Here, if the \enquote{Obtain Admin. Privileges} attack is eventually achieved, the impact of \enquote{Data Leakage} is 3 (\emph{moderate)}, and the impact of \enquote{Denial of Rightful Access to the System} is 2 (\emph{minor)}.


Finally, the Impact and Attack Feasibility attributes are related in a matrix, such as the one shown in Table~\ref{tab: risk matrix}, to determine a Risk attribute, shown as colored squares on the top right of Consequence Nodes, indicating whether the current risk level is acceptable or not.
Here, the Risk Assessment Graph shows a  risk of \emph{significant (S)} \enquote{Data Leakage} and a \emph{moderate (M)} risk of \enquote{Denial of Rightful Access to the System}.

\subsection{Summary of Remarks}\label{sec4: summary of remarks}

In this section, we defined the components required to perform risk assessment using attack graphs:
(1) \emph{consequence nodes},
(2) \emph{attack feasibility attributes}, and
(3) \emph{countermeasure nodes}.
These attributes contribute to extending the threat landscape to a risk landscape.
Apart from the formal definitions, we further showed a practical example of how to utilize this method.
The example is grounded on the risk assessment process described in the DIN VDE V 0831-104.
The described risk assessment process was also used in the project \enquote{Forecast of security requirements and evaluation of possible security concepts for the railway system} provided by the German Center for Rail Traffic Research (DZSF) at the Federal Railway Authority (EBA).

\section{Applicability of Risk Assessment Graphs to Risk Management Standards}\label{sec: applicability of attack graphs to risk management standards}

This section discusses the applicability of Risk Assessment Graphs to ISO/SAE 21434~\cite{21434} and CLC/TS 50701~\cite{50701}, which specify engineering requirements for cybersecurity risk management in the automotive and railway environments, respectively.
We show that our methodology is able to assist different risk analysis methodologies.

\subsection{Risk Assessment According to ISO/SAE 21434}\label{sec: 21434 risk assessment}

\begin{table*}[h]
\renewcommand{\arraystretch}{1.2}
\caption{Example aggregation of attack potential.}
\label{tab: attack potential}
\noindent\makebox[\textwidth]{%
\begin{tabular}{>{\raggedright}p{0.1\textwidth}>{\raggedright}p{0.05\textwidth}|>{\raggedright}p{0.07\textwidth}>{\raggedright}p{0.05\textwidth}|>{\raggedright}p{0.1\textwidth}>{\raggedright}p{0.05\textwidth}|>{\raggedright}p{0.1\textwidth}>{\raggedright}p{0.05\textwidth}|>{\raggedright}p{0.08\textwidth}>{\raggedright}p{0.05\textwidth}}
\toprule
\multicolumn{2}{c}{Elapsed Time} & \multicolumn{2}{c}{\parbox{0.1\textwidth}{Specialist Expertise}}  & \multicolumn{2}{c}{\parbox{0.15\textwidth}{Knowledge of the item (or Component)}}  & \multicolumn{2}{c}{\parbox{0.1\textwidth}{Window of Opportunity}}  & \multicolumn{2}{c}{Equipment}
\tabularnewline
\cmidrule{1-10}
 Enumerate & Value &  Enumerate & Value & Enumerate & Value & Enumerate & Value & Enumerate & Value 
\tabularnewline
\midrule
$<1$ day & 0 & Layman & 0 & Public & 0 & Unlimited & 0 & Standard & 0
\tabularnewline
$<1$ week & 1 & Proficient & 3 & Restricted & 3 & Easy & 1 & Specialized & 4
\tabularnewline
$<1$ month & 4 & Expert & 6 & Confidential & 7 & Moderate & 4 & Bespoke & 7
\tabularnewline
$<6$ months & 17 & Multiple Experts & 8 & Strictly Confidential & 11 & Difficult/None & 10 & Multiple Bespoke & 9
\tabularnewline
$>6$ months & 19 & & & & & & & &
\tabularnewline
\bottomrule
\end{tabular}}
\end{table*}

The ISO/SAE 21434~\cite{21434}, a widely used standard in the automotive domain, defines (1) an \emph{item} as a component or set of components that implement a function at the vehicle level and (2) an \emph{asset} as an object that has value or contributes to value. 
Assets have properties whose compromise may realize \emph{damage scenarios}, which refer to the negative consequences imposed on items.
In this regard, the first step in the risk assessment process, according to ISO/SAE 21434 is the \emph{asset identification}, where the assets of an item are specified, and the possible damage scenarios are evaluated. 
The second step is the \emph{threat scenario identification}, where the potential causes (i.e., threats) of compromise of the assets' properties are analyzed. 
Here, a threat scenario can lead to multiple damage scenarios, and a damage scenario can correspond to multiple threat scenarios. 
The third step is the \emph{impact rating}, where the magnitude of damage or physical harm (i.e impact) from a damage scenario is estimated. 
The damage scenarios are assessed against potential negative consequences, and the impact rating of the damage scenarios is determined to be (1) \emph{negligible}, (2) \emph{moderate}, (3) \emph{major}, or (4) \emph{severe}.
The fourth step is the \emph{attack path analysis}, where threat scenarios are analyzed for identifying attack paths. Here, attack paths are linked to the threat scenarios that can be realized by these attack paths.
The fifth step is the \emph{attack feasibility rating}, where the ease of attack path exploitation is rated as (1) \emph{very low}, (2) \emph{low}, (3) \emph{medium}, or (4) \emph{high}. 
According to the standard, the attack feasibility rating should be based on either (a) an \emph{attack potential}-based approach, (b) a $CVSS^2$-based approach, or (c) an \emph{attack vector}-based approach.
If the rating is based on the attack potential, the attack feasibility should be determined based on core factors including \emph{elapsed time}, \emph{specialist expertise}, \emph{knowledge of the item or component}, \emph{window of opportunity}, and \emph{equipment}.
For each attribute, numerical values can be defined.
The informative parts of the standard propose an example based on the ISO/IEC 18045~\cite{18045}, see~\cref{tab: attack potential}.
The attack potential is then calculated by adding all parameters, the result is then mapped according to~\cref{tab: attack feasibility matrix 21434}.

\begin{table}[h]
\renewcommand{\arraystretch}{1.2}
\caption{Attack Feasibility mapping according to ISO/SAE 21434.}
\label{tab: attack feasibility matrix 21434}
\noindent\makebox[\linewidth]{%
\begin{tabular}[t]{>{\raggedright}p{0.1\textwidth}>{\raggedright}p{0.15\textwidth}}
\toprule
Values & Attack Feasibility
\tabularnewline
\midrule
0-13 & High
\tabularnewline
14-19 & Medium
\tabularnewline
20-24 & Low
\tabularnewline
$>=$25 & Very Low 
\tabularnewline
\bottomrule
\end{tabular}}
\end{table}

The sixth step is the \emph{risk determination}, where the risk of threat scenarios is determined from the impact of the associated damage scenarios and the attack feasibility of the associated attack paths. 
The risk value ranges from 1 (lowest risk) to 5 (highest risk). Here, risk matrices, such as the one shown in Table~\ref{tab: risk 21434}, also proposed by the informative parts of the standard, can also be used for risk determination. 
In addition, if a threat scenario is linked to multiple attack paths, the attack feasibility of these attack paths is aggregated (i.e., the threat scenario is assigned the maximum attack feasibility level of the attack paths). 
The final step in the risk assessment process according to ISO/SAE 21434 is the \emph{risk treatment decision}, where treatment decisions for the identified risks are taken mainly based on the impact and attack feasibility ratings. 
Here, the risk treatment options are (1) \emph{risk avoidance}, by removing risk sources, (2) \emph{risk reduction}, by e.g., inserting countermeasures, (3) \emph{risk sharing or transference}, through e.g., contracts or insurances, or (4) \emph{risk acceptance}, in case of low impact and attack feasibility~\cite{21434}.

\begin{table}[h]
\renewcommand{\arraystretch}{1.2}
\caption{Risk matrix from ISO/SAE 21434~\cite{21434}.}
\label{tab: risk 21434}
\noindent\makebox[\linewidth]{%
\begin{tabular}[t]{>{\raggedright}p{0.12\linewidth}>{\centering}p{0.12\linewidth}>{\centering}p{0.12\linewidth}>{\centering}p{0.12\linewidth}>{\centering\arraybackslash}p{0.06\linewidth}}
\toprule
\multirow{2}[3]{*}{Impact} & \multicolumn{4}{c}{Attack Feasibility}
\tabularnewline
\cmidrule(lr){2-5}
 & Very Low & Low & Medium & High
\tabularnewline
\midrule
Negligible & 1 & 1 & 1 & 1
\tabularnewline
Moderate & 1 & 2 & 2 & 3
\tabularnewline
Major & 1 & 2 & 3 & 4
\tabularnewline
Severe & 2 & 3 & 4 & 5
\tabularnewline
\bottomrule
\end{tabular}}
\end{table}

\subsection{Applicability of Risk Assessment Graphs to ISO/SAE 21434}\label{sec: 21434 attack graphs}

\begin{figure*}[ht]
\includegraphics[width=\textwidth]{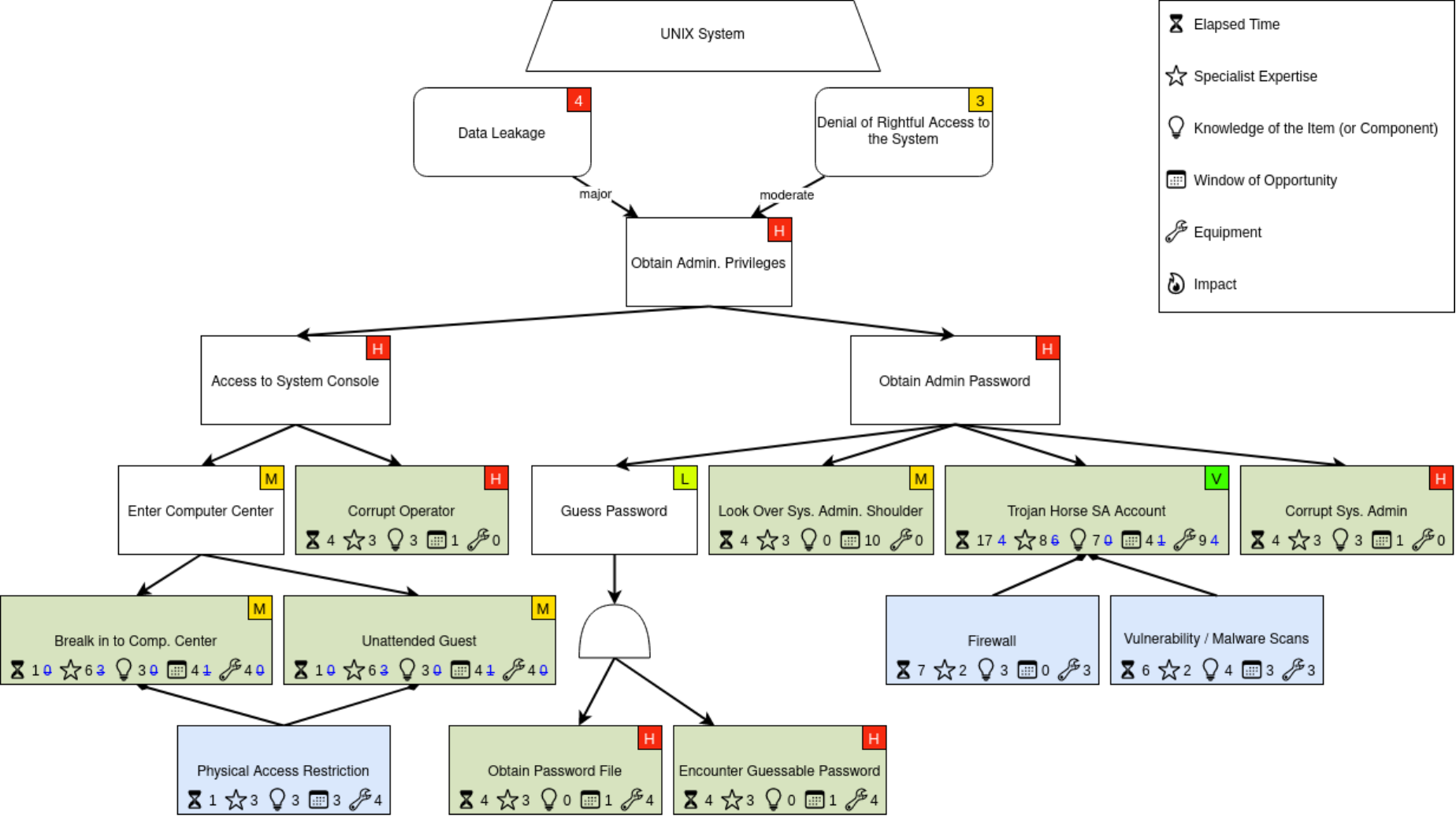}
\caption{Risk graph evaluation of scenario designed by Weiss~\cite{weiss1991} using the risk assessment process described in ISO/SAE 21434~\cite{21434}.}
\label{fig: risk graph 21434}
\end{figure*}

\cref{fig: risk graph 21434} shows an example of a Risk Assessment Graph using ISO/SAE 21434.
The risk assessment process is carried out as follows. First (\emph{asset identification}), in this case, the \emph{item} is a UNIX system, and the \emph{asset} is the administrator privileges. 
Second (\emph{threat scenario identification}), the threat that may compromise the administrator privileges is represented by the topmost Inner Node (i.e., \enquote{Obtain Admin. Privileges}), which leads to two \emph{damage scenarios} (i.e., negative consequences) represented by Consequence Nodes (i.e., \enquote{Data Leakage} and \enquote{Denial of Rightful Access to the System}). 
Third (\emph{impact rating}), the edges relating Consequence Nodes to the topmost Inner Node are assigned with an Impact attribute. 
Here, if the \enquote{Obtain Admin. Privileges} threat is eventually realized, the impact of \enquote{Data Leakage} is \emph{major}, and the impact of \enquote{Denial of Rightful Access to the System} is \emph{moderate}.
Fourth (\emph{attack path analysis}), attack paths are identified through the refinements of Inner Nodes. 
That is, the topmost Inner Node (i.e., \enquote{Obtain Admin. Privileges}) represents the high-level threat (i.e., the main goal of the attack), which is refined into low-level threats (i.e., sub-goals of the attack), represented by successor Inner Nodes until the lowermost Inner Nodes (green nodes) ultimately represent the least significant threats (i.e., elementary attacks).
Hence, all Inner Nodes of the same attack path need to be fulfilled for the \enquote{Obtain Admin. Privileges} threat to be realized (i.e., the main goal of the attack to be achieved). 
Fifth (\emph{attack feasibility rating}), every lowermost Inner Node (green nodes) is assigned with \emph{Elapsed Time} (sandglass), \emph{Specialist Expertise} (star), \emph{Knowledge of the Item (or Component)} (light bulb), \emph{Window of Opportunity} (calendar), and \emph{Equipment} (hammer) attributes, whose value ranges are described in~\cref{tab: attack potential}. 
Then, the Attack Feasibility of every lowermost Inner Node threat is assigned using Table~\ref{tab: attack feasibility matrix 21434}, and predecessor Inner Nodes obtain the attributes from their successor Inner Nodes either conjunctively or disjunctively, as described in Section~\ref{sec: attack graph risk estimation}.

For inner nodes, no specific values are shown as for the risk calculation only the attack feasibility is necessary and if multiple successor nodes have the same attack feasibility, the most critical path is the one which has the lowest sum of attribute values.
Furthermore, the conjunctive calculation is an addition of the values of the successor nodes, resulting in a value of $22$ which according to~\cref{tab: attack feasibility matrix 21434} results in \emph{Low} for the attack feasibility. 
Sixth (\emph{risk determination}), the Impact attribute and the Attack Feasibility of the topmost Inner Node are used to determine the risk of Consequence Nodes. 
Here, using the risk matrix shown in Table~\ref{tab: risk 21434}, the Risk Assessment Graph shows a risk of 4 for \enquote{Data Leakage} and a risk of 3 for \enquote{Denial of Rightful Access to the System}. 
Finally (\emph{risk treatment decision}), the options to \emph{avoid}, \emph{reduce}, or \emph{share/transfer} the identified risks are represented by Countermeasure Nodes. 
Conversely, if the Impact attribute and the Attack Feasibility of the topmost Inner Node are low, and therefore the identified risks are \emph{accepted}, Countermeasure Nodes do not need to be added to the Risk Assessment Graph.


\subsection{Risk Assessment According to CLC/TS 50701}\label{sec: 50701 risk assessment}

Similar to ISO/SAE 21434, CLC/TS 50701 is also describing an \emph{asset-based} risk assessment. 
In this regard, the first step in the risk assessment process according to CLC/TS 50701 is the \emph{system definition}, where the purpose, scope, operational environment, and applicable security standards of the \emph{system under consideration (SUC)} are specified. 
In addition, the valuable objects (i.e., assets) supporting the essential functions of the SuC are identified. 
Assets are classified into \emph{information technology (IT) assets}, whose compromise may lead to business consequences (e.g., loss of revenue), and \emph{operational technology (OT) assets}, whose compromise may lead to physical consequences (e.g., sustained service outages). 
The second step is the \emph{threat landscape}, where a list of potential negative actions or events (i.e., threats) capable of jeopardizing the assets of the SUC is established and maintained.
\begin{figure*}[ht]
\includegraphics[width=\textwidth]{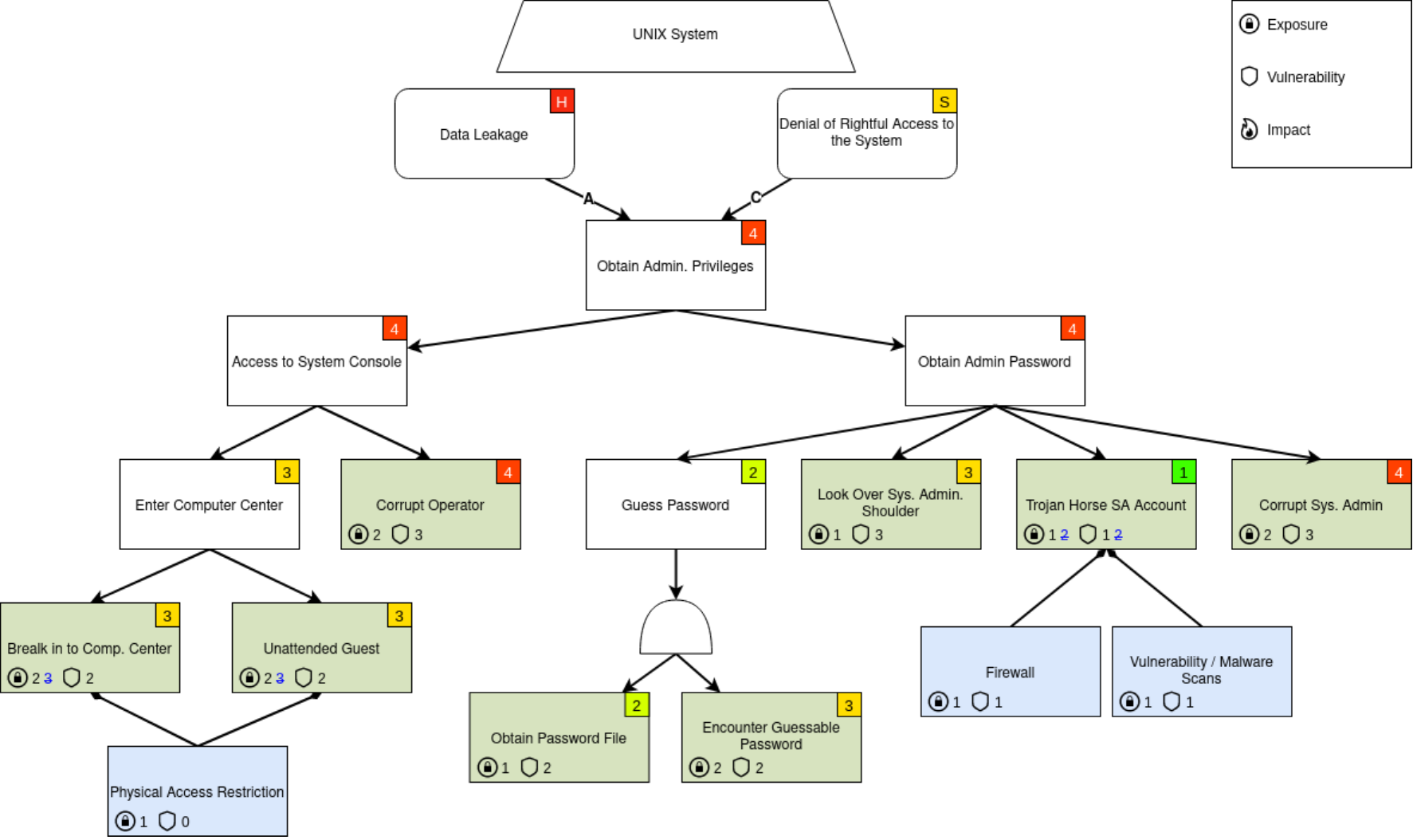}
\caption{Risk graph evaluation of scenario designed by Weiss~\cite{weiss1991} using the risk assessment process described in CLC/TS 50701~\cite{50701}.}
\label{fig: risk graph 50701}
\end{figure*}
Here, threats are classified into internal (i.e., arising from within the system) and external (i.e., arising from without the system), and the skills and motivations driving the threats to exploit the vulnerabilities of the SUC and affect the assets of the SUC are documented.
The third step is the \emph{impact assessment}, where the negative consequences, in terms of \emph{confidentiality, integrity, and availability (CIA)}, imposed on assets are evaluated. 
The CIA impact is assessed qualitatively and ranges from A (highest impact) to D (lowest impact). 
The fourth step is the \emph{likelihood assessment}, where the attack surfaces (i.e., exposure) and the level of expertise and/or resources required to exploit the flaws (i.e., vulnerabilities) of the SUC are evaluated. 
The value of both exposure and vulnerability ranges from 1 (highly restricted logical or physical access for the attacker, vulnerability can only be exploited with high effort) to 3 (easy logical or physical access for the attacker, vulnerability can be exploited with low effort). 
In addition, the likelihood function is $L = EXP + VUL - 1$, and therefore the likelihood of attacks ranges from 1 (highly unlikely) to 5 (highly likely). 
The final step in the risk assessment process, according to CLC/TS 50701, is the \emph{risk evaluation}, where the risk of the documented threats being realized is determined usually by translating the threat landscape into a risk matrix, in which the assessed impact and likelihood of documented threats are related. 
Table~\ref{tab: risk 50701} shows an example of such a risk matrix.

\begin{table*}[h]
\renewcommand{\arraystretch}{1.2}
\caption{Risk matrix from CLC/TS 50701~\cite{50701}.}
\label{tab: risk 50701}
\noindent\makebox[\textwidth]{%
\begin{tabular}[t]{>{\raggedright}p{0.10\linewidth}>{\centering}p{0.12\linewidth}>{\centering}p{0.12\linewidth}>{\centering}p{0.12\linewidth}>{\centering}p{0.12\linewidth}>{\centering\arraybackslash}p{0.12\linewidth}}
\toprule
\multirow{2}[3]{*}{Impact} & \multicolumn{5}{c}{Likelihood}
\tabularnewline
\cmidrule(lr){2-6}
 & 1 & 2 & 3 & 4 & 5
\tabularnewline
\midrule
D & Low & Low & Low & Medium & Significant
\tabularnewline
C & Low & Low & Medium & Significant & High
\tabularnewline
B & Low & Medium & Significant & High & High
\tabularnewline
A & Medium & Significant & High & High & Very High
\tabularnewline
\bottomrule
\end{tabular}}
\end{table*}

\subsection{Applicability of Risk Assessment Graphs to CLC/TS 50701}\label{sec: 50701 attack graphs}

The applicability is demonstrated through the example of a Risk Assessment Graph using CLC/TS 50701, shown in~\cref{fig: risk graph 50701}.
The risk assessment process is carried out as follows. 
First (\emph{system definition}), in this case, the \emph{SUC} is a Linux system. Second (\emph{threat landscape}), the threats (that may jeopardize the assets of the Linux system) are represented by Inner Nodes. 
Similar to ISO/SAE 21434, the topmost Inner Node represents the most significant threat, successor Inner Nodes represent less significant threats, and the lowermost Inner Nodes (green nodes) ultimately represent the least significant threats. 
Hence, the topmost Inner Node is fulfilled when all Inner Nodes of the same attack path are fulfilled. Third (\emph{impact assessment}), the negative consequences are represented by Consequence Nodes (i.e., \enquote{Data Leakage} and \enquote{Denial of Rightful Access to the System}). 
In addition, the edges relating Consequence Nodes to the topmost Inner Node are assigned with an Impact attribute, whose value ranges from A to D. Here, if the \enquote{Obtain Admin. Privileges} threat is eventually realized, the impact of \enquote{Data Leakage} is A, and the impact of \enquote{Denial of Rightful Access to the System} is C. 
Fourth (\emph{likelihood assessment}), every lowermost Inner Node is assigned the Exposure (lock) and Vulnerability (shield) attributes, whose value range from 1 to 3, and the likelihood (red bubble) of every lowermost Inner Node threat is computed using the function $L = EXP + VUL - 1$. 
Then, predecessor Inner Nodes obtain the Exposure and Vulnerability attributes from their successor Inner Nodes either conjunctively or disjunctively, as described in Section~\ref{sec: attack graph risk estimation}.
Finally (\emph{risk evaluation}), the Impact attribute and the likelihood of the topmost Inner Node are used to determine the risk of Consequence Nodes. 
Here, using the risk matrix shown in Table~\ref{tab: risk 50701}, the Risk Assessment Graph shows a high (H) risk of \enquote{Data Leakage} and a significant (S) risk of \enquote{Denial of Rightful Access to the System}.




\section{Evaluation}\label{sec: evaluation}

We identified seven properties that are necessary for a graphical risk assessment process: 
(1) \emph{Attack vectors} to represent the threat landscape.
(2) \emph{Directed acyclic graph} structure due to interdependencies.
(3) \emph{Node attributes} to evaluate the danger that originates from threats.
(4) \emph{Dynamic connectors} due to the complexity and dependencies of attacks/attack steps.
(5) \emph{Edge attributes} for attributes dependent on node connections, like the impact.
(6) \emph{Countermeasure nodes} to represent the existing system and to plan against attacks.
(7) \emph{Consequence nodes} represent what a successful attack impacts.
Section~\ref{sec: related work} discusses that none of the graphical security modeling frameworks from the current literature fulfills all of these seven properties for graphical risk assessment. 
The eight in more detail discussed representatives are again shown in Table~\ref{tab: summary of frameworks}. 
The most relevant framework is the Boolean Logic Driven Markov Processes. 
However, even this one falls short when it gets to attributes, especially edge attributes, and it is not possible to represent consequences either.
Hence, we defined \emph{Risk Assessment Graphs} to fulfill all seven properties. 

\begin{table*}[h]
\rowcolors{2}{gray!10}{gray!40}
\renewcommand{\arraystretch}{1.2}
\caption{Risk Assessment Graphs fulfill all seven identified properties, whereas graphical security modeling frameworks from current literature fulfill at most five properties.}
\label{tab: summary of frameworks}
\noindent\makebox[\textwidth]{%
\begin{tabular}[t]{>{\raggedright}p{0.15\textwidth}>{\raggedright}p{0.06\textwidth}>{\raggedright}p{0.08\textwidth}>{\raggedright}p{0.1\textwidth}>{\raggedright}p{0.09\textwidth}>{\raggedright\arraybackslash}p{0.08\textwidth}>{\raggedright\arraybackslash}p{0.15\textwidth}>{\raggedright\arraybackslash}p{0.1\textwidth}}
\toprule
 & Attack Vectors & DAG Structure & Node Attributes & Dynamic Connectors & Edge Attributes & Countermeasure Nodes & Consequence Nodes
\tabularnewline
\midrule
Attack Trees & \checkmark & - & (\checkmark) & - & - & - & -
\tabularnewline
OWA Trees & \checkmark & - & - & (\checkmark)  & (\checkmark) & - & -
\tabularnewline
Cryptographic DAGs & \checkmark & \checkmark & - & - & - & - & -
\tabularnewline
Bayesian Networks for Security & \checkmark & \checkmark & \checkmark & - & \checkmark & - & -
\tabularnewline
Bayesian Attack Graphs & \checkmark & \checkmark & \checkmark & - & \checkmark & - & -
\tabularnewline
Security Activity Graphs & \checkmark & \checkmark & \checkmark & - & - & \checkmark & -
\tabularnewline
Countermeasure Graphs & \checkmark & \checkmark & \checkmark & - & - & \checkmark & -
\tabularnewline
Boolean Logic Driven Markov Processes & \checkmark & \checkmark & - & (\checkmark) & \checkmark & \checkmark & -
\tabularnewline
Cyber Security Modeling Language & \checkmark & \checkmark & (\checkmark) & - & (\checkmark) & \checkmark & -
\tabularnewline
Risk Assessment Graphs & \checkmark & \checkmark & \checkmark & \checkmark & \checkmark & \checkmark & \checkmark
\tabularnewline
\bottomrule
\end{tabular}}
\end{table*}

Furthermore, we evaluated the \emph{Risk Assessment Graph} method in the project \enquote{Forecast of security requirements and evaluation of possible security concepts for the railway system} provided by the German Center for Rail Traffic Research (DZSF) at the Federal Railway Authority (EBA). 
The project aimed to identify research and standardization demands for security measures of future railway systems. 
In order to achieve this goal, 21 use cases were created, describing how future technologies can be utilized as well as their connections, data flows, and benefits. 
These use cases built the basis for a further security analysis, which was done using Attack Graphs.
First, an Attack Graph for each use case was created to identify the threats and vulnerabilities.
Then, a risk analysis was performed using the method described in~\cref{sec: attack graph risk estimation}.
Next, security measures to mitigate the threats and vulnerabilities were researched and attached to the Risk Assessment Graphs until the risk reached the lowest level.
The security measures were then further investigated to determine the technological level and applicability to the specific railway system as well as their coverage by standards.
So we also wanted to know if the standards that would most likely be applicable for these use cases already require the necessary countermeasures.

During this process, 21 Risk Assessment Graphs with more than 900 Nodes in total (including the Countermeasure Nodes) were created, with the biggest graph containing more than 70 nodes.
It is worth mentioning that if paths were the exact same for different use cases, the attack path was at least once completely evaluated and depicted, whereas it was in other use cases shortened and refers to the other graph.
Furthermore, countermeasures could be used for different attack paths, which resulted in two options for depicting them.
The countermeasure node could have multiple relations to attack paths or the countermeasure node appears multiple times in the same graph.
This is dependent on the specific graph and was decided regarding readability issues.
Additionally, some countermeasure nodes do not contain a single countermeasure, as only the combination of countermeasures would be sufficient to ensure security and can not be evaluated as a stand-alone measure.

Therefore, Risk Assessment Graphs also scale for bigger graphs and for multiple use cases. 
The Risk Assessment Graphs were created using a plugin tool for diagrams.net, which is publicly available on GitHub~\footnote{\label{foot:Plugin}https://incyde-gmbh.github.io/drawio-plugin-attackgraphs/}. 
The Attack Graphs are also publicly available~\footnote{\label{foot: attackgraphs}https://github.com/INCYDE-GmbH/attackgraphs}, and the final versions that include the risk assessment and the countermeasures will also be uploaded publicly in the near future.

Regarding usability, we found the Attack Graphs to be much clearer than alternatives like tables.
Especially using the tool helps as it is also implemented to highlight parent nodes and the attack path with the highest risk.
This makes it easy to follow which path needs to be mitigated to reduce the risk level.
However, countermeasures can defend against multiple attacks, so for bigger graphs duplicating the node is sometimes better than having multiple outgoing edges that cross over the complete graph for better visibility.


\section{Conclusion}\label{sec: conclusion}

This paper presented that the entire process of risk assessment can be performed using a graphical solution, enhancing clarity for all stakeholders and adaptability regarding changed conditions.
We showed that our method enables the replication of the risk assessment process visually and is also compatible with existing risk assessment standards, like the ISO/SAE 21434 or the CLC/TS 50701.
We further verified this method in our project \enquote{Forecast of security requirements and evaluation of possible security concepts for the railway system}, financed by the German Centre of Rail Traffic Research\footnote{https://www.dzsf.bund.de}, by creating 21 graphs of varying sizes in the domain of railway systems.
The Attack Graphs\textsuperscript{\ref{foot: attackgraphs}} are publicly available, as well as the developed open-source plugin\textsuperscript{\ref{foot:Plugin}} for diagrams.net\footnote{https://app.diagrams.net/}, which was used to implement the described method.
Using our method enables a continuous evaluation and reevaluation of the risk landscape, including changing systems, countermeasures, and attacker capabilities.

We identified seven properties necessary to utilize attack graphs for risk assessment: (1) \emph{attack vectors}, (2) \emph{directed acyclic graph structure}, (3) \emph{node attributes}, (4) \emph{dynamic connectors}, (5) \emph{edge attributes}, (6) \emph{countermeasure nodes}, and(7) \emph{consequence nodes}.
These properties were combined to develop the methodology for \emph{Risk Assessment Graphs}, as no existing graphical method combined them.
Risk assessment graphs were further evaluated in the project \enquote{Forecast of security requirements and evaluation of possible security concepts for the railway system} provided by the German Center for Rail Traffic Research (DZSF) at the Federal Railway Authority (EBA).
There, we created 21 graphs of various sizes for future railway systems, using the risk assessment process to identify the need for further research and standardization of countermeasures.

A limitation of a graphical solution is that there is no direct way to include detailed explanations.
Furthermore, especially for bigger graphs, the structure is not as organized as there are a lot of connections.
Resulting in at least some work effort to rearrange the nodes and edges or to double some nodes.
Table-based solutions are usually able to cross-reference data, which should also be able for graph-based solutions, but is not that intuitive and at least in our developed tool not been implemented yet, neither within one graph nor between multiple graphs.

As we move forward, we recognize the potential for further advancements in the field of risk assessment using attack graphs. 
Future research may focus on incorporating real-time threat intelligence feeds and exploring the integration of automation for continuous risk monitoring.

In conclusion, our graphical solution using Risk Assessment Graphs represents a significant step toward a robust and efficient risk assessment.
By harnessing the power of visual representations, we aim to strengthen cybersecurity defenses, ultimately ensuring the resilience and trustworthiness of modern computer systems in the face of evolving threats.



\printbibliography

\end{document}